\documentclass[prb,aps,twocolumn,amsmath,amssymb,floatfix,superscriptaddress]{revtex4-2}
\DeclareMathAlphabet{\mathpzc}{OT1}{pzc}{m}{it}
\DeclareFontFamily{OT1}{pzc}{}
\DeclareFontShape{OT1}{pzc}{m}{it}{ <-> s*[1.1] pzcmi7t }{}
\usepackage{graphicx,amsmath,bm}
\usepackage{float}
\usepackage{subfigure}
\usepackage{mathrsfs}
\usepackage{graphics}
\usepackage{epstopdf}
\usepackage{amssymb}
\usepackage{hyperref}
\usepackage{color}
\usepackage{xcolor}
\usepackage{braket}
\usepackage{multirow}
\usepackage{mathrsfs}
\usepackage{contour}

\begin{document}

\author{Sudin Ganguly}
\email[E-mail: ]{sudinganguly@gmail.com}
\affiliation{Department of Physics, Adamas University, Adamas Knowledge City, Barasat-Barrackpore Road, 24 Parganas North, Kolkata-700 126, India}

\author{Moumita Dey}
\email[E-mail: ]{moumita.dey@adamasuniversity.ac.in}
\affiliation{Department of Physics, Adamas University, Adamas Knowledge City, Barasat-Barrackpore Road, 24 Parganas North, Kolkata-700 126, India}

\author{Santanu K. Maiti}
\email[E-mail: ]{santanu.maiti@isical.ac.in}
\affiliation{Physics and Applied Mathematics Unit, Indian Statistical Institute, 203 Barrackpore Trunk Road, Kolkata-700 108, India}

\title{Spin caloritronics in collinear ferromagnetic helical structures under irradiation}
\begin{abstract}
We study the charge and spin-dependent thermoelectric response of a ferromagnetic helical system irradiated by arbitrarily polarized light, using a tight-binding framework and the Floquet-Bloch formalism. Transport properties for individual spin channels are determined by employing the non-equilibrium Green's function technique, while phonon thermal conductance is evaluated using a mass-spring model with different lead materials. The findings reveal that that light irradiation induces spin-split transmission features, suppresses thermal conductance, and yields favorable spin thermopower and figure of merit (FOM). The spin FOM consistently outperforms its charge counterpart under various light conditions. Moreover, long-range hopping is shown to enhance the spin thermoelectric performance, suggesting a promising strategy for efficient energy conversion in related ferromagnetic systems.
\end{abstract}
\maketitle
\section{\label{sec:intro}Introduction}
Thermoelectric (TE) effects play a central role in modern energy conversion and cooling technologies due to their capacity to convert heat into electrical energy and vice versa~\cite{snyder,tritt,te-review}. While conventional TE devices operate through charge transport, recent developments have highlighted the role of the spin degree of freedom in TE phenomena. This has led to the emergence of spincaloritronics~\cite{wees,boona}, where spin-dependent TE transport enables the integration of spintronics with thermal management, offering new possibilities for efficient and low-power device architectures. In this context, understanding spin-resolved transport behavior is crucial for advancing spin-dependent TE energy conversion. This approach holds promise as a sustainable solution to the global energy crisis by harvesting waste heat.

Traditionally, the efficiency of TE materials is quantified by the dimensionless charge FOM $ZT$, which depends on the temperature and the three transport quantities, electrical conductance, Seebeck coefficient, and thermal conductance. A TE material with $ZT>1$ is usually considered good. However, to match the efficiency of large-scale energy conversion systems, materials need a $ZT$ value around 2 to 3. In the initial stages of TE research, efforts primarily focused on bulk materials, especially semiconductors and their alloys~\cite{semicond1,semicond2,semicond3,semicond4}. However, these systems failed to deliver a significantly high $ZT$. One of the main reasons is that, in bulk materials, electrical and thermal conductivities are linked through the Wiedemann-Franz law~\cite{wf}, which imposes a fundamental limit on improving $ZT$. In their pioneering studies~\cite{hicks1,hicks2}, Hicks and Dresselhaus demonstrated that, in low-dimensional systems, it is possible to independently tune the electronic and thermal conductances. This becomes feasible due to the breakdown of the Wiedemann–Franz law~\cite{kubala}, offering a promising route to achieve enhanced $ZT$. Building on this insight, a wide range of nanoscale systems have been explored to investigate TE behavior by focusing on charge transport~\cite{ms-dress}. These include molecular junctions~\cite{segal,dubi-ventra,reddy,baheti,malen,malen1,nozaki,tan,zerah}, quantum dots~\cite{hoffmann,harman,kuo,wie,trocha}, quantum wires~\cite{hoch,boukai,cornett,bala,qi,chen}, and also biological structures like DNA and proteins~\cite{macia1,macia2,macia3,dong,yli}.

In contrast to conventional charge-based electronic devices, spintronic systems exploit the intrinsic spin of electrons, enabling enhanced operational speed, improved energy efficiency, and miniaturized device architectures, thereby facilitating greater functionality with reduced energy consumption~\cite{zutic}. Beyond these advantages, spin-based TE devices also hold greater promise compared to their charge-based counterparts. Notably, even when the FOM associated with charge transport is relatively low, spin-dependent FOM can still attain higher values. This is a consequence of the FOM scaling with the square of the Seebeck coefficient. While the charge Seebeck coefficient is defined as the average of the spin-resolved Seebeck coefficients, the spin Seebeck coefficient is given by their difference~\cite{uchida-nat,jaworski,eelahi}. Interestingly, in scenarios where the spin-dependent Seebeck coefficients exhibit opposite signs, the total (charge) Seebeck coefficient may become vanishingly small, whereas the spin Seebeck coefficient can be significantly enhanced. This enhancement, in turn, can lead to a markedly improved spin-dependent FOM. Moreover, since a temperature gradient can drive a net spin current in spin-based TE devices, offering the advantage of minimizing thermal losses typically associated with charge transport~\cite{saitoh,sovalen,kimura}. A key distinction of spin-Seebeck devices lies in their unique scaling behavior, which contrasts with traditional charge-driven Seebeck systems. The output power in spin-Seebeck devices depends on the dimension perpendicular to the thermal gradient. Additionally, unlike conventional TE setups where heat and charge currents share the same path, spin-Seebeck configurations allow these currents to travel through different channels. This separation not only provides design flexibility but also opens up possibilities for improving the FOM~\cite{adachi}. Such characteristics have sparked significant interest in the field of spintronics, driving efforts to engineer spin-based TE devices~\cite{kirihara,uchida-jpn,uchida-exprss}.

For the efficient design of spin-based TE devices, a fundamental requirement is the generation of spin-polarized current from an unpolarized electron beam, which essentially demands the separation of up and down spin components~\cite{wolf,zutic}. Various mechanisms have been explored to achieve this goal, such as utilizing ferromagnetic contacts~\cite{johnson,van-son},  spin-orbit coupling~\cite{prem,sudin-jap,niko,foldi}, by applying external magnetic field~\cite{zhang,mucciolo,watson}, etc. However, ferromagnetic contacts often suffer from interface mismatch and low spin injection efficiency~\cite{prinz,schimdt}, while spin-orbit coupling typically requires heavy elements or engineered heterostructures, which can complicate device fabrication and operation~\cite{yhsu}. On the other hand, precise confinement of magnetic fields at the quantum scale is experimentally challenging, thereby limiting their effectiveness in generating spin-polarized currents~\cite{rai}. However, in recent years, chiral systems have received growing attention. The resurgence of interest in this field was primarily driven by the influential study of G\"{o}hler et al.~\cite{gohlar}, which revealed that chiral molecules, owing to their intrinsic structural asymmetry, can function as efficient spin filters. This mechanism of spin-selective transport is known as the chiral-induced spin selectivity (CISS) effect. The CISS effect offers a compelling alternative to the aforementioned methods based on ferromagnetic contacts, spin-orbit coupling, or external magnetic fields. Following this experiment, numerous research groups initiated studies on spin selectivity using a variety of helical molecules and custom-designed helical systems, in an effort to understand the core mechanisms leading to pronounced spin-selective transport~\cite{guo-2012,guo-2014,mishra,guti,trpan}. These cited works employ spin-orbit coupling to achieve spin-filtration both theoretically~\cite{guo-2012,guo-2014,guti,trpan} and experimentally~\cite{gohlar,mishra}. However, weak spin-orbit coupling~\cite{guo-2012} in molecular helices and the need for dephasing~\cite{meth} to observe spin polarization limit the efficiency of CISS, as dephasing suppresses quantum interference essential for optimal response. Interestingly, recent studies showed that CISS can also emerge in achiral systems when exposed to circularly polarized light, without the need for dephasing~\cite{phuc1,phuc2}. In this case, light-matter interaction breaks time reversal symmetry, resulting in spin selective transport, known as Floquet engineered CISS, which differs from the conventional CISS observed in chiral molecules. However, in these scenarios where spin-orbit coupling acts as the primary spin-dependent scattering mechanism, achieving substantial spin polarization remains challenging due to the typically weak coupling strength at the molecular scale~\cite{meth}, and because polarization generally emerges only at specific energy resonances. As an alternative, employing magnetic helical geometries proves highly advantageous, as the spin-splitting effect in such systems is significantly stronger compared to those dominated by spin-orbit interaction~\cite{sarkar}. Furthermore, these systems exhibit spin polarization over a broad energy range, which is particularly appealing from an experimental standpoint.

Building on this foundation, we investigate the spin-dependent TE response of a ferromagnetic helical system irradiated with arbitrarily polarized light. The magnetic configuration is modeled as a uniform collinear ferromagnetic exchange field with all moments aligned along the $z$-quantization axis, providing a clear and tractable framework for analyzing the role of spin polarization. The light-matter interaction offers a mechanism to modulate the degree of spin polarization, thereby enhancing the potential for optimizing TE performance~\cite{ganguly-prb}. The combined effects of intrinsic structural chirality, collinear magnetic ordering, and light-induced modulation establish a robust platform for exploring spin-selective TE behavior.  In a recent work\cite{moumita-prb}, spin polarization has been studied in a magnetic helix in presence of circularly polarized light. In contrast, the present work introduces a generalized formalism that accommodates arbitrary polarization states. This broader approach enables a deeper insight into the influence of light polarization on spin-resolved TE transport and offers improved flexibility for designing spin-driven energy conversion devices.

To investigate the spin-dependent thermoelectric effect, we model the structurally helical ferromagnetic system with a collinear spin configuration within a tight-binding framework. The effect of light irradiation is incorporated through the Floquet-Bloch formalism within the minimal coupling scheme~\cite{sambe,light1,light2}. We employ the non-equilibrium Green's function method combined with the Landauer-B\"{u}ttiker formalism~\cite{datta1,datta2} to calculate spin-resolved electrical conductance, Seebeck coefficient, and electronic thermal conductance. At finite temperatures, phonons contribute to heat transport, and their effect is also included to provide a complete evaluation of both charge and spin FOMs. Our results reveal that {\it light irradiation significantly enhances the overall TE performance of the system, with a particularly strong improvement in the spin-dependent response}.

The rest of the paper is structured as follows. In Sec.~\ref{sec:formalism}, we present the model Hamiltonian for the helical geometry with collinear ferromagnetic ordering, describe the theoretical framework for incorporating arbitrarily polarized light through modified hopping integrals, and outline the formulation of spin-dependent TE quantities along with the phonon contribution to thermal conductance. In the following section (Sec.~\ref{results}), we provide a detailed discussion of the spin-dependent TE responses, both in the absence and presence of light, including the evaluation of charge and spin FOMs. Finally, in Sec.~\ref{conclusion}, we summarize our main findings.
\section{\label{sec:formalism}Theoretical formulation}
\begin{figure}[ht!]
\centering
\includegraphics[width=0.4\textwidth]{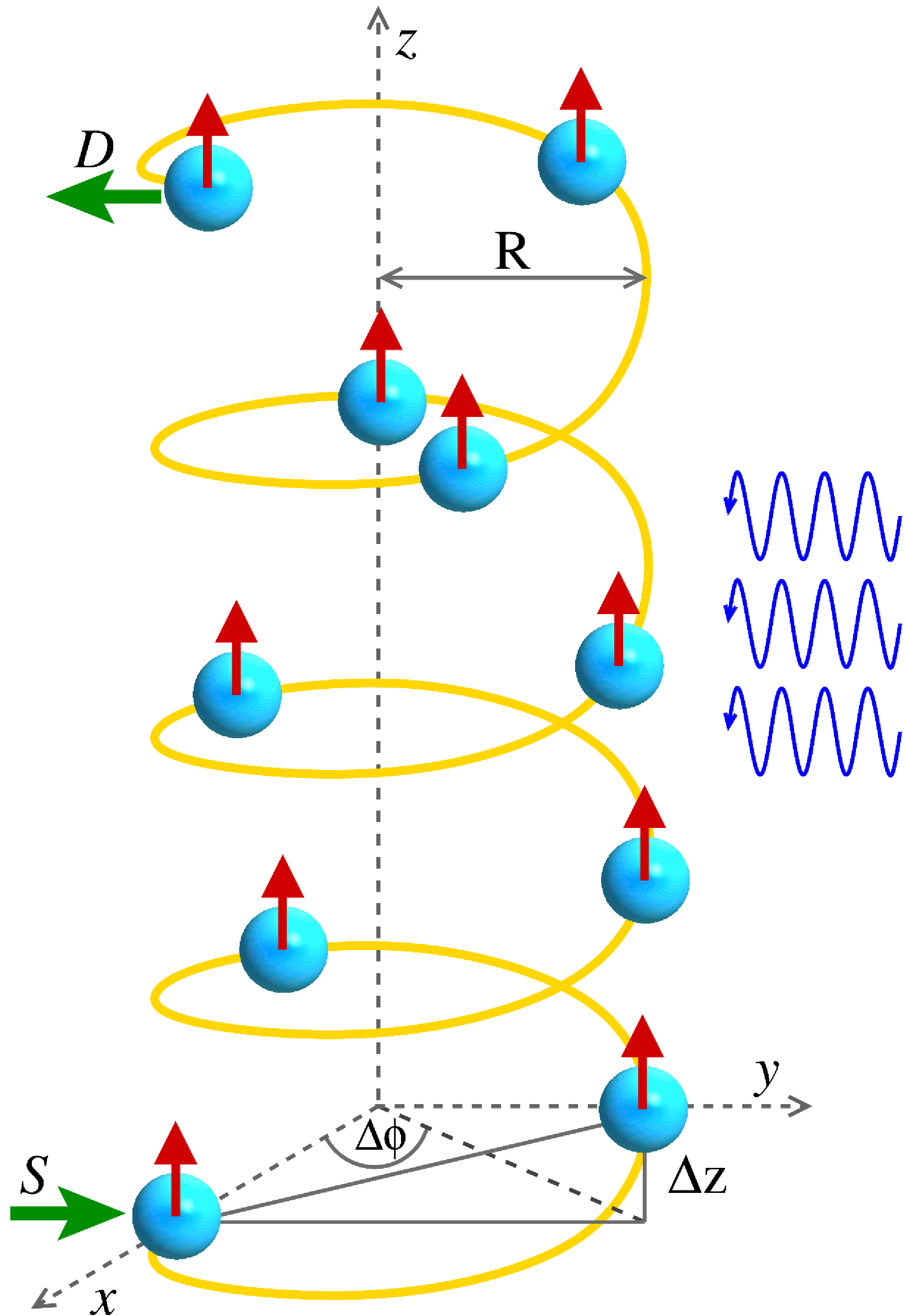}
\caption{(Color online). Schematic representation of an irradiated helical structure incorporating a collinear ferromagnetic spin configuration. The blue balls with the red arrow denoted the magnetic sites. At both ends of the helix, two semi-infinite metallic 1D non-magnetic leads (shown by the green arrows) are attached, referred to as the source ($S$) and drain ($D$). The blue waves represent an arbitrarily polarized light. $R$, $\Delta z$, and $\Delta \phi$ are respectively the radius, stacking distance, and twisting angle of the helix, respectively.}
\label{setup}
\end{figure}

The schematic diagram of a right-handed helix is shown in Fig.~\ref{setup}. The magnetic sites (blue balls) are arranged in a ferromagnetic fashion, with the red up arrows denoting the magnetic moments of the sites. All the moments are identical and assumed be aligned along the $z$-direction which is also chosen as the spin quantized direction in the present work. The end sites of the helix are connected to two 1D semi-infinite, non-magnetic, metallic leads known as the source and drain, labeled as $S$ and $D$ respectively. The ferromagnetic helix is irradiated with an arbitrarily polarized light denoted with the blue waves. The geometry of the helix is defined by the three parameters, namely the radius $R$, the stacking distance $\Delta z$, the vertical height between two nearest-neighbor sites, and 
the twisting angle $\Delta\phi$, the azimuthal angle between two neighboring sites. A large $\Delta z$ indicates that the distance between neighboring sites is significant, allowing us to treat the helix as a short-range helix~\cite{guo-2014}. Conversely, smaller values of $\Delta z$ suggest a long-range helix.

The tight-binding Hamiltonian describing the ferromagnetic helix in the absence of light reads as~\cite{sarkar,moumita-prb}
\begin{eqnarray}
\mathcal{H}_\text{FH} & = & \sum_{ n=1}^N  \mathbf{c}_n^\dagger \left(\bm{\epsilon}_n - \contour[1]{black}{$\mathpzc{h}$}_n \cdot \bm{\sigma} \right) \mathbf{c}_n \nonumber \\
& +& \sum_{n=1}^{N-1} \sum_{m=1}^{N-n} \left( \mathbf{c}_n^\dagger \bm{t}_{nm} \mathbf{c}_{n+m} + h.c. \right),
\label{afm-ham}
\end{eqnarray} 
where $\mathbf{c}_n = \begin{pmatrix}c_{n \uparrow} \\c_{n \downarrow}\end{pmatrix}$ represents the two-component fermioninc annihilation operator  at site $n$ and $\mathbf{c}^\dagger_n$ is the corresponding two-component fermionic creation operator. $\bm{\sigma}$ is the usual Pauli matrices. $\bm{\epsilon}_n$ and $\bm{t}_{nm}$ are the $2 \times 2$ diagonal matrices and are given by
\begin{equation}
\bm{\epsilon}_n= \begin{pmatrix} \epsilon_n & 0\\ 0& \epsilon_n \end{pmatrix} ~~~~\text{and} ~~~ 
\bm{t}_{nm}= \begin{pmatrix} t_{nm} & 0\\ 0& t_{nm} \end{pmatrix} ,
\end{equation}
where $\epsilon_n$ is the on-site energy in the absence of any spin-dependent scattering and $t_{nm}$ represents the hopping amplitude from the site $n$ to $n+m$. The term $\left(\bm{\epsilon}_n -\contour[1]{black}{{$\mathpzc{h}$}}_n \cdot \bm{\sigma}\right)$ is the effective site energy matrix, which represents the interaction between the incoming electron and the local magnetic moments. Here $\contour[1]{black}{{$\mathpzc{h}$}}_n = J\langle \mathbf{S}_n \rangle$ is the spin-dependent scattering (SDS) parameter, where $J$ is the spin-moment exchange interaction and $\langle \mathbf{S}_n \rangle$ is the net spin at the $n$-th magnetic site~\cite{yhsu}.

The hopping amplitude $t_{nm}$ for a helical system can be written as~\cite{av-hop,trpan,guo-2014}
\begin{equation}
t_{nm} = t_1 \mathrm{e}^{\left(l_{nm} - l_1\right)/l_c}
\label{hop}
\end{equation}
where $t_1$ is the nearest-neighbor hopping amplitude. $l_{nm}$ denotes the distance between sites $n$ and $m$, $l_1$ is the nearest-neighbor distance, and $l_c$ is the decay constant. The distance $l_{nm}$ can be written in terms of the geometrical parameters of the helix, i.e., the radius $R$, stacking distance $\Delta z$ and the twisting angle $\Delta \phi$ as~\cite{av-hop,trpan,guo-2014}
\begin{equation}
l_{nm} = \sqrt{4 R^2 \sin{(m\Delta\phi/2)}^2 + (m\Delta z)^2}.
\label{hop-dis}
\end{equation} 

\subsection{Incorporation of light irradiation}
The effect of light irradiation can be incorporate into the helical system in the following way~\cite{light1,light2}. We assume that the propagation direction of the incident light is perpendicular to the axis of the helical system as shown in Fig.~\ref{setup}. When the system is irradiated with light, the corresponding Hamiltonian becomes time dependent. Without loss of generality it can be assumed that any general Hamiltonian is periodic in both space and time under irradiation, that is $H(\vec{x} + \vec{a}, \tau + \mathbb{T}) =  H(\vec{x} , \tau + \mathbb{T}) = H(\vec{x} + \vec{a},  \tau)$, where $\vec{a}$ is the lattice vector and $\mathbb{T}$ is the time period of the incident light and $\mathbb{T}=2\pi/\omega$, $\omega$ being the angular frequency of the incident light. Under this scenario, following the Floquet-Bloch (FB) ansatz, the eigenfunctions can be expressed as
\begin{equation}
\lvert\Psi_{\alpha,\mathbf{k}}(\mathbf{x},\tau)\rangle = \mathrm{e}^{(i\mathbf{k}\cdot\mathbf{x} - i\epsilon_{\alpha,\mathbf{k}}t)} \lvert u_{\alpha,\mathbf{k}}(\mathbf{x},\tau)\rangle,
\end{equation}
where $\epsilon_{\alpha,\mathbf{k}}$ is the quasi-energy corresponding to the $\alpha$th FB state $\lvert u_{\alpha,\mathbf{k}}(\mathbf{x},\tau)\rangle$ and $\mathbf{k}$ is the wave-vector. The FB states are periodic both in $\mathbf{x}$ and $\tau$, and belong to the Hilbert space which is composed of direct product between the space of $\mathbb{T}$-periodic functions and the usual Hilbert space $\mathcal{H}$, that is $\mathcal{S} = \mathcal{H}\otimes\mathbb{T}$. The resulting space is known as the Sambe space~\cite{sambe}. 

To begin with the above stated fact, we consider a general tight-binding Hamiltonian, where there is no spin-dependent interaction, as such interaction will not be affected by the irradiation effect. The Hamiltonian can be written as
\begin{equation}
H = \sum_\alpha \sum_{n,m} t_{nm} c^\dagger_{\alpha,n}(\tau) c_{\alpha,m}(\tau)
\label{ham-ex}
\end{equation}
where $c^\dagger_{\alpha,n}(\tau)$ and $c_{\alpha,n}(\tau)$ are the time-dependent fermionic creation and annihilation operators respectively, and $t_{nm}$ is the hopping integral between sites $n$ and $m$. The Fourier transformations of the creation and annihilation operators are defined as
\begin{eqnarray}
c_{\alpha,\mathbf{k}}(\tau) &=& \frac{1}{\sqrt{N}}\sum_{m} c_{\alpha,\mathbf{k}}(\tau)\mathrm{e}^{i\mathbf{k}\cdot\mathbf{R}_m}\\
c_{\alpha,\mathbf{k}}^\dagger(\tau) &=& \frac{1}{\sqrt{N}}\sum_{m} c_{\alpha,\mathbf{k}}^\dagger(\tau)\mathrm{e}^{-i\mathbf{k}\cdot\mathbf{R}_m}.
\end{eqnarray}

The associated inverse Fourier transforms are given by
\begin{eqnarray}
c_{\alpha,m}(\tau) &=& \frac{1}{\sqrt{N}}\sum_{\mathbf{k}} c_{\alpha,\mathbf{k}}(\tau)\mathrm{e}^{-i\mathbf{k}\cdot\mathbf{R}_m}\\
c_{\alpha,m}^\dagger(\tau) &=& \frac{1}{\sqrt{N}}\sum_{\mathbf{k}} c_{\alpha,\mathbf{k}}^\dagger(\tau)\mathrm{e}^{i\mathbf{k}\cdot\mathbf{R}_m}.
\end{eqnarray}

With the given operators, the Hamiltonian defined in Eq.~\ref{ham-ex} takes the form
\begin{equation}
H = \sum_{\alpha,\mathbf{k}} \sum_{n,m} t_{nm} c^\dagger_{\alpha,\mathbf{k}}(\tau) c_{\alpha,\mathbf{k}}(\tau)\mathrm{e}^{i\mathbf{k}\cdot\left(\mathbf{R}_n - \mathbf{R}_m\right)}.
\end{equation}
The time-periodicity enables the operators to expand as
\begin{eqnarray}
c_{\alpha,\mathbf{k}}(\tau) &=& \sum_p c_{\alpha,\mathbf{k},p}\mathrm{e}^{ip\omega\tau}\\
c^\dagger_{\alpha,\mathbf{k}}(\tau) &=& \sum_p c^\dagger_{\alpha,\mathbf{k},p}\mathrm{e}^{-ip\omega\tau}.
\end{eqnarray}
With the above expanded forms of the operators, the Hamiltonian finally takes the form
\begin{eqnarray}
H_\mathbf{k} &=& \sum_{\alpha,\mathbf{k}} \sum_{n,m} \sum_{p,q} t_{nm} \mathrm{e}^{i\mathbf{k}\cdot\left(\mathbf{R}_n - \mathbf{R}_m\right)} \mathrm{e}^{-i\omega\tau(p-q)} c^\dagger_{\alpha,\mathbf{k},p} c_{\alpha,\mathbf{k},q}\nonumber\\
&=&\sum_{\alpha,\mathbf{k}} \sum_{n,m} \sum_{p,q} \tilde{t}_{nm} \lvert U_{\alpha,\mathbf{k},p}\rangle \mathrm{e}^{-i\omega\tau(p-q)} \langle U_{\alpha,\mathbf{k},q}\rvert,
\end{eqnarray}
where $\tilde{t}_{nm} = t_{nm}\mathrm{e}^{i\mathbf{k}\cdot\left(\mathbf{R}_n - \mathbf{R}_m\right)}$ is the modified hopping integral.

With the composed scalar product and diagonalization of the Hamiltonian $H_\mathbf{k}$, the quasi-energies are obtained as
\begin{eqnarray}
\epsilon_{\alpha,\mathbf{k}} &=& \langle\langle U_{\alpha,\mathbf{k},p} \lvert\mathcal{H}_\mathbf{k}\rvert U_{\alpha,\mathbf{k},p}\rangle\rangle\nonumber\\
&=&\frac{1}{\mathbb{T}}\int_0^\mathbb{T} \langle U_{\alpha,\mathbf{k},p} \lvert\mathcal{H}_\mathbf{k}\rvert U_{\alpha,\mathbf{k},p}\rangle \mathrm{d}\tau,
\end{eqnarray}
where $\mathcal{H}_\mathbf{k} = H_\mathbf{k} - i\hbar\frac{\partial}{\partial\tau}$. Performing the scalar product, we reach to the expression
\begin{eqnarray}
\epsilon_{\alpha,\mathbf{k}} &=& \sum_{n,m}\frac{1}{\mathbb{T}}\int_0^\mathbb{T} \tilde{t}_{nm}\mathrm{e}^{-i\omega\tau(p-q)} \mathrm{d}\tau + q\hbar\omega\delta_{p,q}\nonumber\\
&=&\sum_{n,m} \tilde{t}_{nm}^{p,q} + q\hbar\omega\delta_{p,q},
\label{eff-ham}
\end{eqnarray}
where $q\hbar\omega\delta_{p,q}$ is the Fourier space representation of $-i\hbar\frac{\partial}{\partial\tau}$ and
\begin{equation}
\tilde{t}_{nm}^{p,q} = \sum_{n,m}\frac{1}{\mathbb{T}}\int_0^\mathbb{T} \tilde{t}_{nm}\mathrm{e}^{-i\omega\tau(p-q)} \mathrm{d}\tau.
\end{equation}
Equation~\ref{eff-ham} is the effective time-independent Hamiltonian, where $q\hbar\omega\delta_{p,q}$ represents the site energies and $\tilde{t}_{nm}^{p,q}$ is the hopping integral in the presence of light irradiation. Further, within the minimal coupling scheme, the wave vector transforms as $\mathbf{k}\Rightarrow \mathbf{K}(\tau) = \mathbf{k} + q\mathbf{A}(\tau)/\hbar$, where $q$ denotes the electronic charge and $\mathbf{A}$ the vector potential associated with the incident light. Consequently, the effective hopping integral becomes
\begin{equation}
\tilde{t}_{nm}^{p,q} = \frac{1}{\mathbb{T}}\int_0^\mathbb{T} t_{nm} \mathrm{e}^{i\mathbf{A}\cdot\left(\mathbf{R}_n - \mathbf{R}_m\right)} \mathrm{e}^{-i\omega(p-q)}\mathrm{d}\tau.
\label{eff-hop}
\end{equation}
The modified hopping integral, as defined in Eq.~\ref{eff-hop}, can now be directly incorporated into the helical system under consideration. The term $t_{nm}$ can be identified as the hopping integral as given in Eq.~\ref{hop} and $\left(\mathbf{R}_n - \mathbf{R}_m\right)$ will be replaced by $\left(\mathbf{R}_{nm}- \mathbf{R}_m\right)$, where $\mathbf{R}_{nm}$ and $\mathbf{R}_m$ are the radius vectors of the $(n+m)$th and $m$th sites, respectively. For an arbitrarily polarized light, the vector potential $\mathbf{A}(\tau$) can be expressed as 
\begin{equation}
\mathbf{A}(\tau) = A_x \sin{(\omega \tau)} \boldsymbol{\hat{x}} + A_z \sin{(\omega \tau + \theta)} \boldsymbol{\hat{z}},
\end{equation}
where $A_x$ and $A_z$ are the $x$ and $z$ components, respectively, and $\omega$ is the frequency of the incident light. The light is incident along the $y$ direction.

{\it Calculation of $\left(\mathbf{R}_{nm} - \mathbf{R}_m\right)$}: Assuming the first site of the helix lies on the $x$-axis (see Fig.~\ref{setup}), the components of the radius vector for the $n$th site can be expressed as
\begin{eqnarray}
\left[\mathbf{R}_n\right]_x &=& R\cos{\{(n-1)\Delta\phi\}}\nonumber\\
\left[\mathbf{R}_n\right]_y &=& R\sin{\{(n-1)\Delta\phi\}}\nonumber\\
\left[\mathbf{R}_n\right]_z &=& (n-1)\Delta z.\nonumber
\end{eqnarray}
By the same token, the components of the radius vector of the $(n+m)$th site are
\begin{eqnarray}
\left[\mathbf{R}_{nm}\right]_x &=& R\cos{\{(n+m-1)\Delta\phi\}}\nonumber\\
\left[\mathbf{R}_{nm}\right]_y &=& R\sin{\{(n+m-1)\Delta\phi\}}\nonumber\\
\left[\mathbf{R}_{nm}\right]_z &=& (n+m-1)\Delta z.\nonumber
\end{eqnarray}
Therefore, the different components of the separation vector $\left(\mathbf{R}_{nm} - \mathbf{R}_m\right) = d_x^{nm} \boldsymbol{\hat{x}} + d_y^{nm} \boldsymbol{\hat{y}} + d_z^{nm} \boldsymbol{\hat{z}}$, are given by
\begin{eqnarray}
d_x^{nm} &=& \left[\mathbf{R}_{nm}\right]_x - \left[\mathbf{R}_n\right]_x\nonumber\\
&=& -2R \sin{\left\{\left(n-1+\frac{m}{2}\right)\Delta\phi\right\}} \sin{\left(\frac{m\Delta\phi}{2}\right)}\nonumber\\
d_y^{nm} &=& \left[\mathbf{R}_{nm}\right]_y - \left[\mathbf{R}_n\right]_y\nonumber\\
&=& 2R \sin{\left\{\left(n-1+\frac{m}{2}\right)\Delta\phi\right\}} \sin{\left(\frac{m\Delta\phi}{2}\right)}\nonumber\\
d_z^{nm} &=& \left[\mathbf{R}_{nm}\right]_z - \left[\mathbf{R}_n\right]_z = m\Delta z.
\end{eqnarray}
In this study, the direction of light propagation is chosen to be along the $y$-axis. Given the helical geometry, the incidence angle of light varies along the circumference. As a result, the effective vector potential can be expressed as
\begin{eqnarray}
\mathbf{A}_\text{eff} &=& \boldsymbol{R}_z(\Delta\phi)\mathbf{A}\nonumber\\
&=& \begin{pmatrix}
\cos{\Delta\phi} & \sin{\Delta\phi} & 0 \\
-\sin{\Delta\phi} & \cos{\Delta\phi} & 0 \\
0 & 0 & 1
\end{pmatrix}
\begin{pmatrix}
A_x \sin{(\omega \tau)} \\
0 \\
A_z \sin{(\omega \tau + \theta)}
\end{pmatrix}\nonumber\\
&=& A_x \sin{(\omega \tau)}\cos{(\Delta\phi)}\boldsymbol{\hat{x}} - A_x \sin{(\omega \tau)}\sin{(\Delta\phi)}\boldsymbol{\hat{y}} \nonumber\\
&+& A_z \sin{(\omega \tau)}\boldsymbol{\hat{z}}.\nonumber
\end{eqnarray}
With the expression of $\mathbf{A}_\text{eff}$ and $\left(\mathbf{R}_{nm} - \mathbf{R}_n\right)$, the dot product between them can be computed as
\begin{equation}
\mathbf{A}_\text{eff}\cdot\left(\mathbf{R}_{nm} - \mathbf{R}_n\right) = \Gamma\sin{(\omega\tau + \psi)},
\end{equation}
where
\begin{widetext}
\begin{eqnarray}
\Gamma^2 &=& \left(d_x^{nm} A_x \cos{(\Delta\phi)}\right)^2 + \left(d_y^{nm} A_x \sin{(\Delta\phi)}\right)^2 
+\left(d_z^{nm} A_z \cos{\theta}\right)^2 - d_x^{nm} d_y^{nm} A_x^2 \sin{(2\Delta\phi)} \nonumber\\
&+& 2 d_z^{nm} A_x A_z \left(d_x^{nm}\cos{(\Delta\phi)} - d_y^{nm} \sin{(\Delta\phi)}\right)\\
\psi &=& \arctan{\left[\frac{d_z^{nm} A_z \sin{\theta}}{A_x\left(d_x^{nm}\cos{(\Delta\phi)} - d_y^{nm} \sin{(\Delta\phi)}\right) + d_z^{nm} A_z \cos{\theta}}\right]}.
\end{eqnarray}
\end{widetext}

Finally, with the dot product between $\mathbf{A}_\text{eff}$ and $\left(\mathbf{R}_{nm} - \mathbf{R}_n\right)$, Eq.~\ref{eff-hop} upon integration becomes
\begin{equation}
\tilde{t}_{nm}^{p,q} = t_{nm} J_{p-q}(\Gamma) \mathrm{e}^{i(p-q)\psi},
\label{eff-hop1}
\end{equation}
where $J_{p-q}(\Gamma)$ is the $(p-q)$-th order Bessel function of the first kind. Since the term $\Gamma$ depends on the direction of the bond, the hopping integrals are thus modified according to their direction in the presence of irradiation. 

\subsection{Transmission probability}
To study the TE response, the first key step is to compute the transmission probability, which is achieved using the non-equilibrium Green's function (NEGF) technique. Within this framework, the retarded Green's function is defined as~\cite{datta1,datta2}
\begin{equation}
\mathcal{G}^r = \left[E\mathbb{I}- \mathcal{H}_{\text{FH}}- \Sigma_{S} -\Sigma_{D}\right]^{-1} ,
\end{equation}
where $\Sigma_{S}$ and $ \Sigma_{D}$ represent the contact self-energies of the source and drain, respectively, $\mathbb{I}$ is the $2N \times 2N$ identity matrix. $\mathcal{H}_{\text{FH}}$ is the system Hamiltonian. 

The spin-resolved transmission probability can be expressed using the retarded ($\mathcal{G}^r$) and advanced ($\mathcal{G}^a = \left(\mathcal{G}^r\right)^\dagger$) Green's functions as 
\begin{equation} 
\mathcal{T}_{\sigma \sigma^\prime}= \text{Tr}\left[\Gamma_{\sigma S}~ \mathcal{G}^r~ \Gamma_{\sigma^\prime D} ~\mathcal{G}^a \right] \quad\quad (\sigma, \sigma^\prime \in {\uparrow, \downarrow}). 
\label{eq:trans1} 
\end{equation} 
The term $\mathcal{T}_{\sigma \sigma^\prime}$ represents the transmission probability for an incoming electron with spin $\sigma$ to emerge with spin $\sigma^\prime$. Here, $\Gamma_{\sigma S}$ and $\Gamma_{\sigma^\prime D}$ denote the coupling matrices corresponding to the source and drain leads, characterizing the interaction between the leads and the magnetic helix system.

The net up and down spin transmission probabilities are given by  
\begin{equation} 
\mathcal{T}_\sigma  =  \sum_{\sigma^\prime}\mathcal{T}_{ \sigma^\prime \sigma} , 
\label{spin-trans}
\end{equation} 
where $\sigma, \sigma^\prime \in {\uparrow, \downarrow}$. These spin-dependent transmission functions serve as key ingredients in the evaluation of various TE properties, which are elaborated in the following sub-section.
\subsection{Thermoelectric quantities}  
The spin-dependent TE quantities, namely, the electrical conductance $G_\sigma$, Seebeck coefficient (thermopower) $S_\sigma$, and electronic thermal conductance $k_{\sigma\text{el}}$, can be evaluated using the Landauer's integrals as~\cite{zerah,te-theory2}.
\begin{subequations}
 \begin{eqnarray}
 G_\sigma & = & \frac{e^2}{h} L_{0 \sigma}, \\
 S_\sigma & = & -\frac{1}{e T }\frac{L_{1\sigma}}{ L_{0\sigma}},\\
k_{\sigma \text{el}} & = & \frac{1}{hT}\left( L_{2\sigma} - \frac{L_{1\sigma}^2}{L_{0\sigma}}\right),
 \end{eqnarray}
  \label{eq:TE-quantity}
\end{subequations}

\noindent where spin-resolved Landauer's integral $L_{n\sigma}$ $(n=0,1,2)$ is defined as
\begin{equation}
L_{n\sigma} = - \int   \mathcal{T}_\sigma(E) (E- E_F)^n\frac{\partial f_{\text{FD}}}{\partial E} ~dE, 
\end{equation}
where $h$ represents Planck's constant, $f_{\text{FD}}$ denotes the equilibrium Fermi-Dirac distribution function, and $E_F$ is the Fermi energy. As introduced in Eq.~\ref{spin-trans}, $\mathcal{T}_\sigma(E)$ represents the spin-resolved transmission probability in a two-terminal setup. 

The charge ($c$) and spin ($s$) electrical conductances can be defined in the following way~\cite{te-theory3}
\begin{equation}
  G_{c}  =  G_\uparrow +  G_\downarrow  ~~~ \text{and}~~~  G_{ s} = G_\uparrow -  G_\downarrow .
\end{equation}

The charge and spin Seebeck coefficients are defined as~\cite{te-theory3,te-theory4}
 \begin{equation}
 S_c  =  \frac{1}{2}\left(S_\uparrow + S_\downarrow \right) ~~~\text{and}~~~S_s  =  \left(S_\uparrow - S_\downarrow \right).
\end{equation}

The charge and spin thermal conductances due to electrons are given as~\cite{te-theory3}
 \begin{eqnarray}
 k_{c \text{el}}  =  k_{s \text{el}}= \left(k_\uparrow + k_\downarrow \right).
\end{eqnarray}

The compact expressions for the charge and spin \textit{figures of merit} (FOMs) are given by~\cite{te-theory3}
\begin{equation}
Z_\alpha T = \frac{\lvert G_\alpha\rvert S_{\alpha}^2 ~T}{k_\alpha},
\end{equation}
here, $\alpha ,(=\text{c, s})$ denotes the charge and spin channels, respectively. The total thermal conductance is given by $k_\alpha = k_{\alpha\text{el}} + k_\text{ph}$, where $k_\text{ph}$ represents the phonon contribution. While a thermoelectric FOM of the order of unity is generally considered indicative of a promising response, achieving an economically viable TE performance typically requires $Z_\alpha T \sim 3$~\cite{Tritt-review}. An accurate evaluation of $Z_\alpha T$ necessitates incorporating the phononic component $k_\text{ph}$ in the total thermal conductance. The procedure for calculating $k_\text{ph}$ is detailed in the following sub-section.

\subsection{Calculation of phonon thermal conductance} 
When the temperature difference between the two contact leads approaches zero, the phononic contribution to thermal conductance within the NEGF framework is expressed as~\cite{phonon1,phonon2,phonon3,phonon4} 
\begin{equation} 
k_{\text{ph}}= \frac{\hslash}{2\pi}\int_0^{\omega_c} \mathcal{T}_{\text{ph}}\frac{\partial f_{BE}}{\partial T}\omega d\omega.
\end{equation} 
In this context, $\omega$ refers to the phonon angular frequency, and $\omega_c$ is the upper cutoff frequency for phonons. The calculation assumes that only elastic phonon scattering takes place. The term $f_{BE}$ stands for the Bose-Einstein distribution, while $\mathcal{T}_{\text{ph}}$ represents the phonon transmission function, evaluated using the NEGF approach as 
\begin{equation} 
\mathcal{T}_{\text{ph}} = \text{Tr} \left[ \Gamma_S^{\text{ph}} \mathcal{G}_{\text{ph}} \Gamma_D^{\text{ph}} \left(\mathcal{G}_{\text{ph}}\right)^\dagger \right]. 
\end{equation} 
Here, $\Gamma_{S/D}^{\text{ph}} = i\left[\widetilde{\Sigma}_{S/D} - \widetilde{\Sigma}_{S/D}^\dagger\right]$ characterizes the thermal broadening from the source and drain phonon reservoirs, with $\widetilde{\Sigma}_{S/D}$ denoting the self-energy matrices associated with the respective contacts. The phononic retarded Green's function for the Ferromagnetic helix structure is defined by 
\begin{equation} 
\mathcal{G}_{\text{ph}} = \left[ {\mathbb M} \omega^2 - {\mathbb K} - \widetilde{\Sigma}_S - \widetilde{\Sigma}_D \right]^{-1}. 
\end{equation} 
In this formulation, ${\mathbb M}$ is a diagonal mass matrix, while ${\mathbb K}$ encodes the spring constants of the helical structure. The explicit forms of these matrices are given by
\begin{eqnarray}
{\mathbb M} = \begin{pmatrix}
M_{11} & 0 & \cdots & 0 \\
0 & M_{22} & 0 & \vdots \\
\vdots & 0 & \ddots & \vdots \\
0 & \cdots & \cdots & M_{NN}
\end{pmatrix},\\
{\mathbb K} = \begin{pmatrix}
K_{11} & K_{12} & \cdots & K_{1N} \\
K_{21} & K_{22} & K_{23} & \vdots \\
\vdots & K_{32} & \ddots & \vdots \\
K_{N1} & \cdots & \cdots & K_{NN}
\end{pmatrix},
\end{eqnarray}
where each diagonal entry $M_{nn}$ represents the mass of the $n$-th atom. In the spring-constant matrix ${\mathbb K}$, the diagonal elements $K_{nn}$ correspond to self-restoring forces, while the off-diagonal elements $K_{nm}$ describe the coupling between atoms $n$ and $m$. 

The matrix ${\mathbb K}$ captures the vibrational dynamics of the helical structure as follows. As discussed in the electronic Hamiltonian, the helical structure exhibits hopping beyond nearest-neighbors, with the hopping integrals scaled as $t_{nm} = t_1 \mathrm{e}^{\left(l_{nm} - l_1\right)/l_c}$ (see Eqs.~\ref{hop} and \ref{hop-dis}). Similarly, to model phonon vibrations in such structures, one must include couplings beyond the nearest-neighbors, represented by the spring constants $K_{nm}$. These spring constants are scaled analogously to the electronic hopping integrals, $K_{nm} = K_1\mathrm{e}^{\left(l_{nm} - l_1\right)/l_c}$, where $K_1$ is the nearest-neighbor spring constant, and the remaining symbols have the same meanings as in Eqs.~\ref{hop} and \ref{hop-dis}. Once all the spring constants are determined, the self-restoring terms $K_{nn}$ can be obtained by summing the elements of the $n$-th row of the spring-constant matrix ${\mathbb K}$. For example, $K_{11}$ is obtained by summing the first row of ${\mathbb K}$, $K_{22}$ from the second row, and so on. The nearest-neighbor spring constant $K_1$ depends on the specific material under consideration and can be evaluated as discussed below.

The self-energy terms $\widetilde{\Sigma}_S$ and $\widetilde{\Sigma}_D$ are of the same dimension as ${\mathbb M}$ and ${\mathbb K}$, and are obtained from the expression $\Sigma_{S/D} = -K_{S/D} \exp\left[2i\sin^{-1}\left(\omega/\omega_c\right)\right]$, where $K_{S/D}$ stands for the spring constant at the interface between the lead and the helical region.

These spring constants are computed by taking the second derivative of the interatomic potential proposed by Harrison~\cite{harrison}. For one-dimensional leads, which lack transverse modes~\cite{kittel}, the spring constant simplifies to $K_{S/D} = 3dc_{11}/16$. In contrast, for a three-dimensional structure such as the helix, the spring constant is given by $K_1 = 3d(c_{11} + 2c_{12})/16$, where $d$ denotes the bond length and $c_{11}$, $c_{12}$ are the relevant elastic moduli. The cutoff frequency for phonons in the 1D lead is determined by the relation $\omega_c = 2\sqrt{K_{S/D}/M}$, with $M$ being the atomic mass and $K_{S/D}$ is the spring constant of the lead.

\section{\label{results}Numerical results and discussion}
Before presenting the results, let us first clarify that the present work focuses exclusively on right-handed helices. Throughout the paper, energy is measured in electron-volts (eV). The on-site energies $\epsilon_{n\uparrow}$ and $\epsilon_{n\downarrow}$ in the FM helix are set to zero, and the nearest-neighbor hopping (NNH) strength is chosen as $t_1 = 1\,$eV. For the leads, we adopt $\epsilon_0 = 0$ and $t_0 = 2.5\,$eV. To operate in the wide-band limit~\cite{wide-band}, we ensure $t_0 > t_1$. The motivation for considering the wide-band limit is to avoid missing any propagating energy channels of the helix that is sandwiched between the leads. Accordingly, the energy window of the leads, given by $\epsilon_0 - 2t_0 \leq E \leq \epsilon_0 + 2t_0$, must fully encompass the allowed energy range of the helix system. It is important to note that the choice $\epsilon_0 = 0$ is not unique. Any other value satisfying the above condition is equally valid. The coupling strengths between the helix and the source and drain leads, denoted by $\tau_S$ and $\tau_D$, are both set to $0.8\,$eV. We have verified through extensive numerical calculations that for other choices of tight-binding parameters (excluding the coupling constants), the qualitative features of the results remain unchanged. The vector potential is expressed in the unit of $el_1/\hbar$. Unless stated otherwise, the structural parameters of the helix are fixed as follows: radius $R = 2.5\,$\AA, twisting angle $\Delta\phi = 5\pi/9$, stacking distance $\Delta z = 1.5\,$\AA, and the decay constant $l_c=0.9\,$\AA. These choices of helix parameters give rise to a long-range hopping scenario~\cite{guo-2014}. In our model, the magnetic moments of the FM helix are taken to be equal in magnitude and oriented along the positive $z$-direction, with $\mathpzc{h}_n = \mathpzc{h} = 0.25$. 

In scenarios where the driving frequency is relatively low, periodic perturbations induce non-negligible interactions between the central system and multiple auxiliary configurations that emerge due to the time dependence. These auxiliary structures, often referred to as replica states, reflect the fact that time-periodic systems can be reformulated as higher-dimensional static analogs, with one additional dimension introduced via the temporal periodicity~\cite{light1}. In this regime, the influence of higher-order Fourier components-indexed by integers $p$ and $q$ (see Eq.~\ref{eff-hop1}), becomes prominent, thereby complicating the  energy structure of the system. Two significant issues arise: first, the enlargement of the effective Hilbert space tends to reduce the average level spacing, often pushing it below thermal energy scales. This limits the ability to resolve discrete energy levels. Second, the expansion of the effective system size can surpass the spin coherence length~\cite{spin-length}, reducing the likelihood of sustaining coherent spin-dependent transport across the system. These constraints collectively diminish the effectiveness of low-frequency modulation in controlling spin-resolved electronic behavior.

To avoid these limitations, we focus on a regime where the frequency of the applied radiation is significantly higher than the intrinsic electronic hopping scale, specifically satisfying $\omega \gg 4t_1$. In this high-frequency domain, the temporal modulation does not couple different sidebands effectively, allowing the dynamics to be accurately described by only the central (zeroth-order) Floquet mode. This simplification leads to a reduced effective model where the system behaves similarly to its undriven counterpart, but with modified hopping amplitudes.

The key modification introduced by the radiation appears through a scaling factor involving the zeroth-order Bessel function, $J_0(\Lambda)$, where $\Lambda$ serves as a parameter characterizing the nature of the incident light. This factor encapsulates how the periodic field reshapes the tunneling processes, thereby influencing transport phenomena in a non-trivial yet controllable way.

In our setup, we select a driving frequency in the vicinity of $10^{16}$ Hz, placing it within the ultraviolet or extreme-ultraviolet spectral range. The radiation intensity is set around $10^{11}\,$W/cm$^2$, which remains well within current experimental capabilities. Indeed, numerous prior investigations have successfully explored regimes involving even more intense fields~\cite{cwd1,cwd2}, affirming the physical viability and safety of the parameters chosen in our study.

It should be noted here that in real materials, photoexcitation can modify chemical bonding and the band structure through bond softening, structural distortions, and many-body renormalization. These effects are not included in our model. This approximation is justified because the driving frequency used here is about $10^{16}\,$Hz, whereas typical optical phonon frequencies are about $10^{12}\,$Hz. Since the ions cannot respond on such fast timescales, the irradiation primarily affects only the electronic sector through Floquet-band formation and Peierls-phase-induced hopping renormalization, while lattice distortions and electron-phonon coupling can be safely neglected. We note that the corresponding field intensity is $\sim 10^{11}\,\text{W/cm}^2$, which may cause heating in real materials, but this can be mitigated in engineered systems~\cite{light2,artificial} with larger effective lattice constants that allow lower driving frequencies and reduced field strengths.

We begin our study with Fig.~\ref{trans}, which illustrates the spin-resolved transmission probabilities as a function of energy, both in the absence and presence of light irradiation. The number of sites in the helix fixed at $N=20$. The transmission spectra for spin-up and spin-down electrons are represented by black and red curves, respectively. In the absence of light (Fig.~\ref{trans}(a)), 
\begin{figure}[ht!]
\centering
\includegraphics[width=0.4\textwidth]{fig2a.eps} \vskip 0.1 in
\includegraphics[width=0.4\textwidth]{fig2b.eps} 
\caption{(Color online). Spin-resolved transmission probability $\mathcal{T}_\sigma$ $(\sigma = \uparrow, \downarrow)$ as a function of energy in the (a) absence and (b) presence of light. The light parameters are $A_x = 0.3$, $A_z = 0.2$, and $\theta = 0$. The number of sites in the helix is $N = 20$, with the SDS parameter set to $\mathpzc{h} = 0.25$. The physical parameters of the ferromagnetic helix are $R = 2.5\,$\AA, $\Delta\phi = 5\pi/9$, $\Delta z = 1.5\,$\AA, and $l_c = 0.9\,$\AA. The region marked by the blue ellipse in panel (b) is magnified in the inset to highlight the behavior near $E = 0$. Black and red curves represent the up and down spin channels, respectively.}
\label{trans}
\end{figure}
\begin{figure*}[ht!]
\centering
\includegraphics[width=0.33\textwidth,height=0.21\textwidth]{fig3a.eps}\hfill\includegraphics[width=0.33\textwidth,height=0.21\textwidth]{fig3b.eps}\hfill\includegraphics[width=0.33\textwidth,height=0.21\textwidth]{fig3c.eps}\vskip 0.1 in
\includegraphics[width=0.33\textwidth,height=0.21\textwidth]{fig3d.eps}\hfill\includegraphics[width=0.33\textwidth,height=0.21\textwidth]{fig3e.eps}\hfill\includegraphics[width=0.33\textwidth,height=0.21\textwidth]{fig3f.eps}
\caption{(Color online). Behavior of various thermoelectric quantities as a function of Fermi energy at room temperature ($T = 300\,$K). The upper panels [(a)-(c)] correspond to the absence of light, while the lower panels [(d)-(f)] show the results in the presence of light. Panels (a) and (d) display the electrical conductance ($G_\alpha$), (b) and (e) show the thermopower ($S_\alpha$), and (c) and (f) present the electronic thermal conductance ($\kappa_\text{el}$). All system parameters are the same as those used in Fig.~\ref{trans}. The subscript $\alpha$ denotes charge ($c$) and spin ($s$) components, represented by blue and green curves, respectively.}
\label{gsk}
\end{figure*}
a finite mismatch is observed between the spin-up and spin-down channels. This asymmetry arises from the spin-dependent scattering induced by the FM helical ordering in the system. An additional noteworthy feature is the non-uniform spacing between the transmission resonances. At lower energies, the peaks are densely clustered, while at higher energies, they become more widely spaced. Such behavior is typical in systems with extended hopping~\cite{moumita-prb}, as is the case in our helical model. The lack of uniformity in peak separation is a manifestation of broken electron-hole symmetry. As the range of electron hopping increases, the energy gaps between successive resonant states become larger in the high-energy regime, leading to the observed broadening in peak separation. In Fig.~\ref{trans}(b), we show the spin-resolved transmission spectrum in the presence of linearly polarized light, with the light parameters set to $A_x = 0.3$, $A_z = 0.2$, and $\theta = 0$. This configuration corresponds to linear polarization in the $x$-$z$ plane. An interesting feature emerges near zero energy. While in the absence of light, both spin-up and spin-down transmissions remain finite in this region, upon introducing light, the down spin transmission is significantly suppressed around $E = 0$, indicating the opening of a light-induced spin-dependent gap. This behavior is highlighted by a blue ellipse in Fig.~\ref{trans}(b), with a magnified view shown in the inset. The inset reveals a noteworthy phenomenon where the spin-up and spin-down transmission channels cross each other. While spin-splitting is a fundamental requirement for realizing spin-based FOM, such transmission crossings serve as an equally important secondary condition for enhancing spin-dependent thermoelectric performance~\cite{ganguly-prb}. These crossings introduce sharp energy gradients in the spin-resolved transmission functions, which can lead to a significant enhancement in the spin thermopower, as will be discussed in the following sections. A similar crossing between the spin-up and spin-down transmission channels appears near $E = 2.7\,$eV in the absence of light (Fig.~\ref{trans}(a)), which would in principle yield a strong thermoelectric response if the Fermi energy were set close to this value. However, under typical experimental conditions, the Fermi level is tuned near the band center, where the density of extended states is highest and transport measurements are most reliable. In contrast, approaching the band edges, such as around $2.7\,$eV, reduces the availability of extended states and increases sensitivity to disorder, rendering this regime less suitable for experimental probing. For this reason, our analysis in this present work focuses on achieving favorable thermoelectric performance near the band center.

For completeness, the average spin-resolved density of states (DOS) corresponding to Fig.~\ref{trans} is provided in the Supplementary Material (Fig.~\ref{s2}). The average spin-resolved DOS profiles are fully consistent with the features observed in the transmission spectra of Fig.~\ref{trans}.

We now turn our attention to the analysis of various TE quantities, including electrical conductance, thermopower, thermal conductance, and the FOM, evaluated at room temperature ($T = 300\,$K). Both charge and spin-dependent TE responses are investigated for systems with long-range helical order. A comprehensive discussion is presented by systematically varying key physical parameters over a broad range and examining the effects of different light polarizations. This approach allows us to assess the robustness and reliability of the observed TE behavior under diverse conditions. 

TE properties are plotted against the Fermi energy in Fig.~\ref{gsk}. The top panel illustrates the scenario without light exposure, whereas the bottom panel depicts results under linearly
 polarized light. All system and light parameters used here are identical to those specified in Fig.~\ref{trans}.

The variation of electrical conductance $G_\alpha$ (in units of $e^2/h$) with Fermi energy $E_F$ is shown in Figs.~\ref{gsk}(a) and (d), where $\alpha$ refers to either charge or spin. The charge conductance is depicted in blue, while the spin conductance is represented in green. The charge conductance $G_c$ oscillates smoothly with Fermi energy, remaining mostly between $1$ and $2$ (in units of $e^2/h$). An oscillatory pattern is also observed in the spin conductance $G_s$, but with a smaller amplitude and fluctuations around zero. These oscillations reflect resonant features originating from the energy-dependent transmission channels, as seen in Fig.~\ref{trans}(a).

In the presence of light, $G_c$ is suppressed, particularly in the low-energy region (Fig.~\ref{gsk}(d)). A pronounced reduction appears near $E_F = 0$, consistent with the transmission gap induced by light, as shown in Fig.~\ref{trans}(b). However, no appreciable measurable change is observed in the spin electrical conductance under illumination. The reason is as follows. The individual spin-resolved conductances ($G_\uparrow$ and $G_\downarrow$) both decrease in the presence of light, as illustrated in Fig.~\ref{s1} of the Supplementary Material. This reduction naturally leads to a visible suppression in the charge conductance, since $G_c = G_\uparrow + G_\downarrow$. Yet, because the decreases in $G_\uparrow$ and $G_\downarrow$ are nearly identical, the spin electrical conductance $G_s = G_\uparrow - G_\downarrow$ remains largely unaffected, thereby showing no appreciable difference between the illuminated and light-free cases. Moreover, within the considered Fermi-energy window, $G_s$ takes both positive and negative values. A positive (negative) $G_s$ indicates that the spin-up (spin-down) contribution dominates. This behavior is fully consistent with the spin-resolved electrical conductances presented in Fig.~\ref{s1} in the Supplementary Material.

Since FOM scales with the square of the thermopower ($S$), achieving high $S$ values is crucial for enhancing TE performance. In spin-dependent TE systems, the thermopower contributions from spin-up and spin-down electrons can carry opposite signs. When this occurs, their algebraic combination may yield an amplified overall spin FOM. For this selecting an appropriate Fermi energy ($E_F$) is important where such spin-dependent asymmetries arise.

To observe thermopower with opposite spin contributions, one must identify a narrow energy region where the transmission function becomes asymmetric around $E_F$. Importantly, the spin-up and spin-down transmission curves should exhibit opposite slopes near this energy to generate differing TE responses.

The thermopower is obtained from Eq.~\ref{eq:TE-quantity}(b), which involves the first-order Landauer moment $L_1$. In this expression, the transmission function is weighted by the factor $(E - E_F)\frac{\partial f_\text{FD}}{\partial E}$, where $f_\text{FD}$ is the Fermi-Dirac distribution. This product acts as a thermally broadened, antisymmetric window centered at $E_F$. If the transmission function $\mathcal{T}(E)$ is symmetric about $E_F$, the integral vanishes, leading to zero thermopower-regardless of the transmission magnitude~\cite{sudin-super}.

However, if $\mathcal{T}(E)$ exhibits a notable asymmetry, particularly where spin-up and spin-down components have opposite slopes~\cite{ganguly-prb}, the integral becomes finite and large in magnitude. This asymmetry is key to generating substantial and sign-resolved spin thermopower, thereby improving the overall spin TE efficiency.

In the absence of light (Fig.~\ref{gsk}(b)), both $S_c$ and $S_s$ show oscillatory behavior with moderate amplitude, typically within the range of $\pm 100\,\mu$V/K. The spin thermopower $S_s$ exhibits more pronounced fluctuations due to the difference in spin-resolved transmission spectra. In this regime, the asymmetry in the transmission functions is relatively weak, which limits the magnitude of the Seebeck response and therefore leads to low values of the charge thermopower. For the spin channel, although a crossing between the spin-up and spin-down transmissions appears near $E = 0.5\,$eV, the associated asymmetry remains modest. As a result, the spin Seebeck coefficient attains only a moderate value around $E_F = 0.5\,$eV. Under light irradiation (Fig.~\ref{gsk}(e)), the thermopower exhibits a significant enhancement. The charge thermopower attains a peak value of nearly 300$\,\mu$V/K, while the spin thermopower reaches as high as 600$\,\mu$V/K. This pronounced increase in spin thermopower arises from the crossing of the spin-up and spin-down transmission spectra, as discussed earlier in connection with Fig.~\ref{trans}(b). These crossings lead to transmission profiles with opposite slopes for the two spin channels around certain Fermi energies, thereby producing thermopower contributions of opposite signs. The algebraic combination of these contributions results in large values of $S_s$. This behavior highlights the crucial role played by the interplay between spin splitting and spin-channel crossings in amplifying the spin-dependent Seebeck effect, a feature that becomes accessible only under the influence of light.

Figures~\ref{gsk}(c) and \ref{gsk}(f) depict how the electronic thermal conductance $k_e$ (in pW/K) varies with the Fermi energy $E_F$, for the cases without and with light, respectively. In the absence of light (Fig.~\ref{gsk}(c)), $k_e$ displays a relatively high and oscillatory behavior, with values ranging roughly between 200 and 400$\,$pW/K across the considered Fermi energy window. This trend arises from the combined effect of resonant features in the transmission spectrum and thermal broadening.

Upon introducing light (Fig.~\ref{gsk}(f)), $k_e$ is significantly suppressed across most of the Fermi energy range, dropping below 100$\,$pW/K in several regions, especially around $E_F = 0$, $k_e$ is only a few pW/K. This notable decrease in thermal conductance arises from the suppression of electronic transmission due to light irradiation, consistent with the patterns seen in the corresponding conductance and transmission spectra. Since the FOM is inversely proportional to $k_e$, this considerable reduction in electronic thermal conductance under light irradiation is particularly advantageous. It enhances the overall TE performance by boosting the FOM, making the system more efficient for energy conversion.

The systems considered in this study are relatively small in size, typically extending over just a few nanometers. Consequently, the phonon contribution to thermal conductance is typically expected to be smaller than its electronic counterpart. However, since the analysis is carried out at room temperature, the phononic contribution becomes non-negligible and must be accounted for in the accurate evaluation of the FOM, which we discuss in the following discussion.

Within the tight-binding framework, electronic properties are determined by onsite energies and hopping amplitudes, which can be suitably scaled and treated in a largely material-independent manner. In contrast, phononic transport explicitly depends on material-specific parameters such as atomic mass, interatomic spacing, and spring constants. Consequently, the constituent materials must be specified for both the leads and the central helical system in the phonon calculations to ensure physical realism. In this study, the helical molecule is therefore modeled using carbon atoms, representative of organic helices, while silicon and germanium leads are chosen to provide vibrational reservoirs with distinct phonon spectra, thereby avoiding trivial spectral matching and enabling a clearer assessment of helix-induced transport effects. Similar choices of lead materials have been adopted and discussed in several previous studies~\cite{phonon1,phonon2,phonon3}.

The spring constants of the germanium and silicon leads are $13.71\,\text{N/m}$ and $16.87\,\text{N/m}$, respectively~\cite{phonon3,harrison}. With the reported values of $c_{11}$ and $c_{12}$ for single-crystal benzene~\cite{hasel} and considering the interatomic distance as $l_1$ used in the electronic case, for the central helical molecule, the nearest-neighbor spring constant comes out to be around $K_1=10.17\,$N/m. At the interface, we assume the bonding occurs between two distinct atomic species, one from the lead material and the other from the helical structure. To estimate the effective coupling at this junction, we average the respective spring constants and atomic masses of the two regions. This allows us to determine the cut-off phonon frequencies relevant to each configuration. We point out that the transverse phonon contributions are omitted in our 1D lead setup. However, due to the three-dimensional geometry of the helix, such transverse modes are naturally present within the molecular structure. These additional vibrational degrees of freedom in the helix contribute to the phonon-mediated thermal conductance and must be considered in a complete TE analysis. With the chosen parameter values, the phonon cut-off frequencies for the germanium and silicon leads are found to be $26.2$ and $40.2\,$Trad/sec, respectively. 

Figures~\ref{phonon}(a) and (b) depict the phonon transmission coefficient $\mathcal{T}_{\text{ph}}$ as a function of phonon frequency $\omega$ for silicon and germanium leads, respectively. For the silicon leads, the transmission spectrum exhibits pronounced Fabry-P\'{e}rot-like oscillations~\cite{phonon3}, 
\begin{figure}[ht!]
\centering
\includegraphics[width=0.24\textwidth,height=0.19\textwidth]{fig4a.eps}\hfill\includegraphics[width=0.24\textwidth,height=0.19\textwidth]{fig4b.eps}
\vskip 0.1 in
\includegraphics[width=0.24\textwidth,height=0.19\textwidth]{fig4c.eps}
\caption{(Color online). Phonon transmission probability $\mathcal{T}_{\text{ph}}$ as a function of phonon angular frequency $\omega$ for (a) silicon, and (b) germanium leads. (c) Phonon thermal conductance $k_{\text{ph}}$ as a function of temperature $T$ for the same set of leads. Black and red curves represent results for  silicon and germanium leads, respectively. The central system is composed of carbon in all cases.}
\label{phonon}
\end{figure}
arising from phonon wave interference between interfaces that reflect resonant vibrational modes confined within the finite-sized central region. Such oscillatory behavior, however, is less prominent for the germanium leads due to their larger atomic mass. In the case of silicon (Fig.~\ref{phonon}(a)), the transmission remains relatively high and smooth over a broad frequency range, followed by a sharp decline at the cut-off frequency $\omega_c = 40.2\,$Trad/sec. In contrast, the germanium case (Fig.~\ref{phonon}(b)) exhibits weaker oscillations with broader peak spacing and a lower cut-off at $\omega_c = 26.2\,$Trad/sec. Overall, the silicon lead (Fig.~\ref{phonon}(a)) supports phonon transport over a broader frequency range, where the Fabry-P\'{e}rot-like transmission peaks become progressively closer and sharper with increasing frequency. These differences stem from the distinct elastic constants and atomic masses of the two materials, which strongly influence the interference pattern and phonon transport characteristics across the junction.

Figure~\ref{phonon}(c) illustrates the phonon thermal conductance $k_{\text{ph}}$ as a function of temperature $T$ for the two types of leads discussed above. In both cases, $k_{\text{ph}}$ increases gradually at low temperatures and tends to saturate as the temperature rises, reflecting the enhanced thermal excitation of phonon modes at higher energies. Across the entire temperature range, the silicon lead exhibits a higher thermal conductance than the germanium lead. This behavior can be attributed to the combined influence of two key material parameters, namely, the spring constant and the atomic mass. A larger spring constant, as in silicon, increases the phonon cut-off frequency, thereby enhancing phonon transport. Conversely, a heavier atomic mass, as in germanium, lowers the vibrational frequencies and suppresses thermal transport. Although germanium possesses a relatively large spring constant, its higher atomic mass shifts the vibrational spectrum toward lower frequencies, resulting in a smaller $k_{\text{ph}}$. Silicon, on the other hand, benefits from both a high spring constant and an intermediate atomic mass, enabling it to support a broader range of phonon frequencies and thus a higher phonon thermal conductance. At room temperature, the phonon thermal conductance $k_{\text{ph}}$ in Fig.~\ref{phonon}(c) attains distinct values for the two materials: approximately $55\,$pW/K for the silicon lead and about $37\,$pW/K for the germanium lead.

Having computed all the TE quantities, namely, the electrical conductance, thermopower, and the thermal conductances due to both electrons and phonons, we now present the behavior of the charge and spin FOMs at room temperature, which encapsulates their combined effect. The results are illustrated in Fig.~\ref{zt-fom}, where 
\begin{figure}[ht!]
\includegraphics[width=0.24\textwidth]{fig5a.eps}\hfill\includegraphics[width=0.24\textwidth]{fig5b.eps}\vskip 0.1 in
\includegraphics[width=0.24\textwidth]{fig5c.eps}\hfill\includegraphics[width=0.24\textwidth]{fig5d.eps}
\caption{(Color online). Variation of charge and spin figures of merit, $Z_cT$ and $Z_sT$, with Fermi energy at room temperature ($T = 300\,$K). The upper panel corresponds to the absence of light, while the lower panel shows the results in the presence of light. All system parameters are the same as those in Fig.~\ref{trans}. The results correspond to systems with silicon and germanium leads, represented by black and red curves, respectively, highlighting the phononic contributions.}
\label{zt-fom}
\end{figure}
the upper panel shows the system response in the absence of light, and the lower panel corresponds to the case under light irradiation. The black and red curves represent the results for silicon and germanium leads, respectively, as discussed earlier in the context of phonon thermal conductance. Since the phonon thermal conductance of silicon leads is comparatively higher than that of germanium leads, systems with germanium leads consistently exhibit higher FOMs than those with silicon leads. In the absence of light (Figs.~\ref{zt-fom}(a) and (b)), both the charge ($Z_cT$) and spin ($Z_sT$) FOMs remain below unity for both the leads, indicating poor TE efficiency. However, upon light irradiation (Figs.~\ref{zt-fom}(c) and (d)), a notable enhancement is observed in both quantities. The maximum $Z_cT$ reaches approximately unity (Fig.~\ref{zt-fom}(c)), while $Z_sT$ peaks close to 2.5 around $E_F = 0$ for the germanium lead (Fig.~\ref{zt-fom}(d)). For both the leads, $Z_sT$ consistently exceeds unity under light, signifying favorable spin TE performance. Overall, light irradiation enhances the charge TE response for specific leads and universally boosts the spin TE performance, regardless of the lead material.

As it is established that the spin FOM ($Z_sT$) exhibits a more favorable response in the presence of light for a particular set of light parameters, we next investigate whether this enhanced performance persists under different light configurations. We begin by studying the effect of the phase of polarization $\theta$, as shown in Fig.~\ref{fom-lights}(a). The driving field components are set to $A_x = 0.3$ and $A_z = 0.2$, with three phase values $\theta = 0$, $\pi/4$, and $\pi$ shown by black, red, and green curves, respectively. Here, the phononic contribution is 
\begin{figure}[ht!]
\centering
\includegraphics[width=0.24\textwidth]{fig6a.eps}\hfill\includegraphics[width=0.24\textwidth]{fig6b.eps}
\caption{(Color online). Behavior of the spin figure of merit $Z_sT$ as a function of Fermi energy under different light parameters. (a) Effect of light polarization angle $\theta$ with fixed driving field components $A_x = 0.3$ and $A_z = 0.2$, for $\theta = 0$, $\pi/4$, and $\pi/2$, represented by black, red, and green curves, respectively. (b) Effect of different light polarizations: linear ($A_x = 0.3$, $A_z = 0$, $\theta = 0$; black curve), circular ($A_x = A_z = 0.2$, $\theta = \pi/2$; red curve), and elliptical ($A_x = 0.2$, $A_z = 0.3$, $\theta = \pi/4$; green curve). All other system parameters are the same as those used in Fig.~\ref{trans}.}
\label{fom-lights}
\end{figure}
taken from the germanium lead. Remarkably, $Z_sT$ reaches a value of approximately to 7 around $E_F = -0.5$  for $\theta = \pi/2$ (green curve) and about the same $Z_sT$ value for $\theta=\pi/4$ (red curve) near $E_F=0$. For $\theta=0$, the spin FOM about 2, indicating strong spin-dependent TE performance. Next, we examine the effect of light polarization type on $Z_sT$, as depicted in Fig.~\ref{fom-lights}(b). We consider three distinct cases: linearly, circularly, and elliptically polarized light. For linearly polarized light, the parameters are $A_x = 0.3$, $A_z = 0$, and $\theta = 0$. For circular polarization, we set $A_x = A_z = 0.2$ and $\theta = \pi/2$. For elliptical polarization, the values are $A_x = 0.2$, $A_z = 0.3$, and $\theta = \pi/4$. In all cases, $Z_sT$ exceeds unity, confirming favorable spin TE response. Among them, linearly polarized light yields the highest $Z_sT$, reaching about 3.5, followed by circular polarization with a maximum near 2, and elliptical polarization just exceeding unity. As mentioned earlier, the light-induced transmission gap, including the crossing between the up- and down-spin transmission channels, arises from the combined effects of Floquet-band hybridization and Peierls-phase–induced hopping renormalization. As seen in Eq.~\ref{eff-hop1}, the renormalized hopping amplitudes depend explicitly on the driving-field parameters $A_x$, $A_z$, and $\theta$. Consequently, all three components of the driving field collectively tune the band-structure modification and directly influence the resulting thermoelectric response.

For completeness, we have provided the corresponding charge and spin electrical conductances, charge and spin thermopowers, and the electronic thermal conductance for all sets of light parameters used in Fig.~\ref{fom-lights} in the Supplementary Material (Figs.~\ref{s3}--\ref{s7}). These additional plots fully support the analysis and observations presented in Fig.~\ref{gsk}, and the trends observed therein are consistently reflected in the spin FOM shown in Fig.~\ref{fom-lights}.

So far, all the results have been discussed for a long-range FM helix, where the term long-range refers to hopping connections that extend beyond the nearest-neighbor sites. To better understand the influence of the helix structure itself on the TE response, particularly the spin FOM, it is important to explore how $Z_sT$ varies with the hopping range of the helix. A suitable geometric parameter to characterize this range is the decay constant $l_c$. By varying $l_c$ from a very small value to a large one, the system smoothly evolves from a short-range helix, dominated by nearest-neighbor hopping, to a long-range helix with significant higher-order hopping amplitudes. In particular, the limiting case of $l_c = 0$ corresponds to purely nearest-neighbor hopping. 

To provide a clearer picture of how higher-order hopping emerges in the system, we plot the hopping amplitudes $t_i$ (for $i > 1$), scaled by the nearest-neighbor hopping $t_1$, as a function of the decay constant $l_c$, as shown in Fig.~\ref{fom-lc}(a). The hopping amplitudes are evaluated using Eqs.~\ref{hop} and \ref{hop-dis}, while all other geometric parameters of the helix are kept fixed, as specified in Fig.~\ref{trans}. The plot includes hopping contributions from the second- to sixth-nearest neighbors. As expected, for $l_c = 0$ and values close to zero, all higher-order hopping amplitudes vanish, indicating a strictly nearest-neighbor system. As $l_c$ increases, the second-neighbor hopping becomes finite first. Around $l_c = 1\,$\AA, hopping amplitudes up to the fourth neighbor become significant, and with further increase in $l_c$, even more distant hopping terms begin to contribute.

Using this information, we now plot the maximum $Z_sT$ as a function of $l_c$, as shown in Fig.~\ref{fom-lc}(b). 
The maximum value is obtained by scanning $Z_sT$ over a Fermi energy window ranging from $E_F = -1$ to $E_F = 1$. Here, we consider three different sets of light parameters: $(A_x = 0.3, A_z = 0.2, \theta = 0)$, $(A_x = 0.2, A_z = 0.3, \theta = 0)$, and $(A_x = 0.3, A_z = 0.2, \theta = \pi/4)$, denoted by black, red, and blue curves, 
\begin{figure}[ht!]
\centering
\includegraphics[width=0.24\textwidth]{fig7a.eps}\hfill\includegraphics[width=0.24\textwidth]{fig7b.eps}
\caption{(Color online). (a) Higher-order hopping amplitude $t_i$ ($i=2,3,4,5,6$) as a function of $l_c$. The amplitudes are scaled with the neareat-neighbor hopping $t_1$. (b) $Z_sT$ as a function of $l_c$ for different light parameters: $(A_x=0.3, A_z=0.2, \theta=0)$ (black with circle symbol), $(A_x=0.2, A_z=0.3, \theta=0)$ (red with square symbol), and $(A_x=0.3, A_z=0.2, \theta=\pi/4)$ (blue with triangle symbol). All other physical parameters of the FM helix are the same as Fig.~\ref{trans}.}
\label{fom-lc}
\end{figure}
respectively. For small values of $l_c$ (near 0), $Z_sT$ is  below unity. As $l_c$ increases, all three curves show a 
notable enhancement in $Z_sT$, highlighting the growing role of higher-order hopping processes. The black curve (circle symbols) exhibits a pronounced peak at $l_c \approx 0.7\,$\AA, where $Z_sT$ reaches a maximum of about 12, suggesting an optimal decay length for maximizing spin TE efficiency. Additional peaks appear around $l_c = 2.5\,$\AA\ and $l_c = 3.5\,$\AA, with $Z_sT$ values of approximately 8 and 7, respectively. The red curve (square symbols), corresponding to reversed $A_x$ and $A_z$ values, shows a consistently favorable response, with $Z_sT$ remaining above unity across a broad range of $l_c$. The blue curve (triangle symbols) exhibits a peak $Z_sT$ of around 27 near $l_c = 0.7\,$\AA, with an addition peak about 11 near $l_c = 2.5\,$\AA. These observations underscore the intricate interplay between system geometry, light-matter interaction, and higher-order hopping in determining spin transport properties in helical systems. Overall, short-range helical configurations are not promising for spin TE applications, while long-range helices generally yield significantly better performance. Importantly, the spin FOM can be effectively tuned by adjusting the decay constant $l_c$.

We have also investigated how the decay constant $l_c$, which encodes the effective range of the helical vibrational couplings, governs phonon transport. In particular, we analyzed the influence of $l_c$ (and thus the underlying helical geometry) on both the phonon transmission spectra and the resulting phonon thermal conductance. These results, presented in the Supplementary Material with detailed description and analysis, clearly demonstrate that the phonon transport is highly sensitive to the helical geometry. To maintain clarity and focus in the main manuscript, these extended phonon-transport results are provided exclusively in the Supplementary Material. For a comprehensive discussion, please refer to Figs.~\ref{s88}--\ref{s9} therein.

{\bf Existing helical structure with ferromagnetic ordering:} Experimental realizations of structural helices that retain essentially collinear ferromagnetic ordering have been reported in several studies and provide direct physical motivation for the geometry used in our work. For example, electrodeposited Co-Fe nanosprings display a clear helical morphology while exhibiting soft, largely collinear ferromagnetic behavior as evidenced by hysteresis and magnetic force microscopy measurements \cite{nam2017}. Helical metallic nanostructures fabricated by glancing-angle deposition (GLAD) likewise preserve bulk-like ferromagnetic alignment while their curvature and torsion significantly influence coercivity and magnetization reversal modes \cite{kesapragada2006,GLADBook}. More recently, tomography-based experimental studies combined with micromagnetic simulations have shown that three-dimensional magnetic nanohelices can exhibit magnetization reversal mechanisms strongly controlled by geometry (curvature/torsion), while the local magnetization largely follows a uniform axis rather than forming a continuously rotating spin spiral \cite{fullerton2024}. Together, these works demonstrate that a helical structural backbone with collinearly aligned magnetic moments is experimentally attainable and that the geometry alone can induce substantial modifications to magnetic and spin-transport properties, which are directly relevant to the ferromagnetic ordering with helical structure considered in our study.

\section{\label{conclusion}Closing remarks}
In this work, we have investigated the charge and spin-dependent TE response of a ferromagnetic helical system irradiated by arbitrarily polarized light. The system is modeled using a tight-binding framework, where the light-matter interaction is incorporated via the Floquet-Bloch formalism under the minimal coupling scheme. A detailed derivation is performed to obtain the effective hopping integrals in the presence of light with arbitrary polarization.

TE quantities such as spin-resolved electrical conductance, thermopower, and electron-mediated thermal conductance are calculated using spin-dependent transmission probabilities derived via the non-equilibrium Green’s function (NEGF) method. Phonon thermal conductance is computed through a mass-spring model using NEGF, considering different materials (carbon, silicon, and germanium) for the leads, while the central system is assumed to be carbon-based. The effects of light irradiation under various input conditions on the charge and spin FOMs are analyzed at room temperature.

Our key findings are summarized below:

    $\bullet$ The spin-resolved transmission spectra show clear separation between up and down spins in the ferromagnetic helix.

    $\bullet$ Light irradiation leads to a suppression in transmission for both spin channels and induces crossings, which are crucial for enhancing the spin thermopower.

    $\bullet$ Light suppresses $G_\uparrow$ and $G_\downarrow$, reducing the charge conductance, while the spin conductance changes in profile but not appreciably in magnitude.

    $\bullet$ The spin thermopower consistently exceeds the charge thermopower under light exposure.

    $\bullet$ Electron-mediated thermal conductance remains very low under illumination.

    $\bullet$ Among the lead materials considered, germanium yields the lowest phonon thermal conductance at room temperature.

    $\bullet$ The spin FOM is found to be consistently more favorable than its charge counterpart.

    $\bullet$ Favorable spin FOM values exceeding unity are obtained for a wide range of light polarizations.

    $\bullet$ Systems with long-range hopping exhibit enhanced spin TE performance, highlighting the importance of higher-order hopping processes.

To the best of our knowledge, spin-dependent TE effects in ferromagnetic helical systems under arbitrarily polarized light have not been addressed in existing literature. The insights and physical interpretations presented here may serve as a useful foundation for future studies. Our results also suggest a promising direction for achieving efficient energy conversion using similarly structured irradiated ferromagnetic materials.

\section*{Supporting Information}
Average spin-resolved density of states as a function of energy, spin-resolved electrical conductance, charge and spin thermoelectric quantities, namely charge and spin electrical conductances and Seebeck coefficients, charge electronic thermal conductance as a function of Fermi energy under different light parameters, phonon transmission probability as a function of phonon angular frequency for different decay constant $l_c$, phonon thermal conductance as a function of $l_c$.

\bibliographystyle{apsrev4-2}


\pagebreak
\widetext
\begin{center}
\textbf{\large Supplemental Materials: Spin caloritronics in collinear ferromagnetic helical structures under irradiation}
\end{center}
\setcounter{equation}{0}
\setcounter{figure}{0}
\setcounter{table}{0}
\setcounter{section}{0}
\makeatletter
\renewcommand{\theequation}{S\arabic{equation}}
\renewcommand{\thefigure}{S\arabic{figure}}
\renewcommand{\thesection}{S\arabic{section}}
\renewcommand{\bibnumfmt}[1]{[S#1]}
\renewcommand{\citenumfont}[1]{S#1}
\section{Spin-resolved density of states}
For a system with spin degrees of freedom, the retarded Green's function is block diagonal in spin space:
\begin{equation}
{\mathcal G}^r(E) = 
\begin{pmatrix}
{\mathcal G}^r_{\uparrow\uparrow}(E) & {\mathcal G}^r_{\uparrow\downarrow}(E)\\
{\mathcal G}^r_{\downarrow\uparrow}(E) & {\mathcal G}^r_{\downarrow\downarrow}(E)
\end{pmatrix}
\end{equation}

The average spin-resolved density of states (DOS) for spin $\sigma=\uparrow,\downarrow$ is given by
\begin{equation}
\rho_\sigma(E) = -\dfrac{1}{N\pi} \text{Im Tr}\left[{\mathcal G}^r_{\sigma\sigma}(E)\right],
\end{equation}
where $N$ is the number of sites in the helix.

With the above definition, we plot the average spin-resolved DOS $\rho_\sigma$ as a function of energy in Fig.~\ref{s2}, both in the absence and presence of light. The DOS profiles are fully consistent with their corresponding transmission spectra. In Fig.~\ref{s2}(b), we also highlight the crossing between the up- and down-spin DOS in the presence of light, which mirrors the behavior discussed in Fig.~2 of the main text.

\begin{figure}[H]
\centering
\includegraphics[width=0.45\textwidth]{dos_no_light.eps}\hfill\includegraphics[width=0.45\textwidth]{dos_light.eps}
\caption{(Color online.) Average spin-resolved density of states $\rho_\sigma$ as a function of the Fermi energy in the (a) absence and (b) presence of light. The light parameters are $A_x = 0.3$, $A_z = 0.2$, and $\theta = 0$. The number of sites in the helix is $N = 20$, with the SDS parameter set to $\mathpzc{h} = 0.25$. The physical parameters of the ferromagnetic helix are $R = 2.5\,$\AA, $\Delta\phi = 5\pi/9$, $\Delta z = 1.5\,$\AA, and $l_c = 0.9\,$\AA. The region marked by the blue ellipse in panel (b) is magnified in the inset to highlight the behavior near $E = 0$. Black and red curves represent the up and down spin channels, respectively.}
\label{s2}
\end{figure} 

\section{Spin-resolved electrical condductance}
Figure~\ref{s1} illustrates the behavior of $G_\uparrow$ and $G_\downarrow$ as a function of the Fermi energy, both in the absence and in the presence of light. The occurrence of positive and negative values in the spin electrical conductance follows directly from the definition $G_s = G_\uparrow - G_\downarrow$, as evidenced by Fig.~\ref{s1}. In contrast, the charge electrical conductance remains strictly positive, consistent with its definition $G_c = G_\uparrow + G_\downarrow$. The trends observed in Fig.~\ref{s1} are in clear agreement with those presented in Figs.~3(a) and (d)  of the main text.
\begin{figure}[H]
\centering
\includegraphics[width=0.45\textwidth]{gu_gd.eps}\hfill\includegraphics[width=0.45\textwidth]{gu_gd_light.eps}
\caption{(Color online.) Spin-resolved electrical conductance $G_\sigma$ $(\sigma = \uparrow, \downarrow)$ as a function of the Fermi energy in the (a) absence and (b) presence of light. The light parameters are $A_x = 0.3$, $A_z = 0.2$, and $\theta = 0$. The number of sites in the helix is $N = 20$, with the SDS parameter set to $\mathpzc{h} = 0.25$. The physical parameters of the ferromagnetic helix are $R = 2.5\,$\AA, $\Delta\phi = 5\pi/9$, $\Delta z = 1.5\,$\AA, and $l_c = 0.9\,$\AA. Black and red curves represent the up and down spin channels, respectively.}
\label{s1}
\end{figure}

\section{Thermoelectric quantities under different light parameters}
The different thermoelectric quantities, namely the charge and spin electrical conductances, charge and spin thermopowers, and the electronic thermal conductance, for all sets of light parameters used in Fig.~6 of the manuscript are plotted below. The analysis and observations presented in Fig.~3 are in full agreement with all the plots shown here, and their impact is clearly reflected in Fig.~6, where the charge and spin figures of merit (FOMs) are displayed. Detailed descriptions of each plot are provided in the respective captions.
\begin{figure}[H]
\centering
\includegraphics[width=0.33\textwidth]{g_ax0.3_ay0.2_phi_pib4_ge.eps}\hfill\includegraphics[width=0.33\textwidth]{s_ax0.3_ay0.2_phi_pib4_ge.eps}\hfill\includegraphics[width=0.33\textwidth]{ke_ax0.3_ay0.2_phi_pib4_ge.eps}
\caption{(Color online.) Behavior of charge and spin thermoelectric quantities as a function of Fermi energy at room temperature ($T = 300\,$K). (a) electrical conductance ($G_\alpha$), (b) thermopower ($S_\alpha$), and (c) electronic thermal conductance ($k_\text{el}$).  The subscript $\alpha$ denotes charge ($c$) and spin ($s$) components, represented by blue and green curves, respectively. The light parameters are $A_x=0.3$, $A_y=0.2$, $\phi=\pi/4$. All system parameters are the same as those used in Fig.~2 in the main text.}
\label{s3}
\end{figure} 
\begin{figure}[H]
\centering
\includegraphics[width=0.33\textwidth]{g_ax0.3_ay0.2_phi_pib2_ge.eps}\hfill\includegraphics[width=0.33\textwidth]{s_ax0.3_ay0.2_phi_pib2_ge.eps}\hfill\includegraphics[width=0.33\textwidth]{ke_ax0.3_ay0.2_phi_pib2_ge.eps}
\caption{(Color online.) Same as Fig.~\ref{s3} with light parameters $A_x=0.3$, $A_y=0.2$, $\phi=\pi/2$.}
\label{s4}
\end{figure} 
\begin{figure}[H]
\centering
\includegraphics[width=0.33\textwidth]{g_ax0.3.eps}\hfill\includegraphics[width=0.33\textwidth]{s_ax0.3.eps}\hfill\includegraphics[width=0.33\textwidth]{ke_ax0.3.eps}
\caption{(Color online.) Same as Fig.~\ref{s3} with light parameters $A_x=0.3$, $A_y=0$, $\phi=0$.}
\label{s5}
\end{figure} 
\begin{figure}[H]
\centering
\includegraphics[width=0.33\textwidth]{g_ax0.2_ay0.3_phi_pib4.eps}\hfill\includegraphics[width=0.33\textwidth]{s_ax0.2_ay0.3_phi_pib4.eps}\hfill\includegraphics[width=0.33\textwidth]{ke_ax0.2_ay0.3_phi_pib4.eps}
\caption{(Color online.) Same as Fig.~\ref{s3} with light parameters $A_x=0.2$, $A_y=0.3$, $\phi=\pi/4$.}
\label{s6}
\end{figure} 
\begin{figure}[H]
\centering
\includegraphics[width=0.33\textwidth]{g_ax0.2_ay0.3_phi_pib2.eps}\hfill\includegraphics[width=0.33\textwidth]{s_ax0.2_ay0.3_phi_pib2.eps}\hfill\includegraphics[width=0.33\textwidth]{ke_ax0.2_ay0.3_phi_pib2.eps}
\caption{(Color online.) Same as Fig.~\ref{s3} with light parameters $A_x=0.2$, $A_y=0.3$, $\phi=\pi/2$.}
\label{s7}
\end{figure}

\section{Phonon transport: Role of $l_c$}
The phonon transmission probabilities and the corresponding phonon thermal conductance are computed using the formalism described in the manuscript. The central scattering region is modeled as a helical structure composed of carbon atoms. The leads are treated as one-dimensional for simplicity. In the main text, we consider two different types of leads: silicon and germanium. For both cases, we evaluate the phonon transmission probability as a function of the phonon frequency for various values of the decay constant $l_c$. By tuning $l_c$, the helical geometry effectively transitions from a short-range to a long-range helix, allowing the influence of higher-order vibrational couplings to be clearly manifested in the resulting transmission characteristics.

\begin{figure}[H]
\includegraphics[width=0.33\textwidth]{si_trans_ph_lc0.1.eps}\hfill\includegraphics[width=0.33\textwidth]{si_trans_ph_lc0.5.eps}\hfill\includegraphics[width=0.33\textwidth]{si_trans_ph_lc1.eps}\vskip 0.1 in
\includegraphics[width=0.33\textwidth]{si_trans_ph_lc2.eps}\hfill\includegraphics[width=0.33\textwidth]{si_trans_ph_lc3.eps}\hfill\includegraphics[width=0.33\textwidth]{si_trans_ph_lc4.eps}
\caption{(Color online). Phonon transmission probability $\mathcal{T}_{\text{ph}}$ as a function of phonon angular frequency $\omega$ for silicon leads. (a) $l_c=0.1\,$\AA, (b) $l_c=0.5\,$\AA, (c) $l_c=1\,$\AA, (d) $l_c=2\,$\AA, (e) $l_c=3\,$\AA, and (f) $l_c=4\,$\AA. The temperature is fixed at $T=300\,$K.}
\label{s88}
\end{figure} 
In Fig.~\ref{s88} we plot the phonon transmission probability $\mathcal{T}_{\rm ph}(\omega)$ 
for the helical molecule with silicon leads for several values of the decay length 
$l_c$. For the shortest decay length, $l_c=0.1$ (Fig.~\ref{s88}(a)), the spectrum is densely 
oscillatory (Fabry-P\'erot-like), with many closely spaced resonances extending across most 
of the silicon phonon window, which reflects strong interference between rapidly decaying 
vibrational components. At $l_c=0.5$ (Fig.~\ref{s88}(b)) the fine oscillatory structure begins to smooth out, although multiple resonance and anti-resonance features remain. Increasing to $l_c=1$ (Fig.~\ref{s88}(c))
further broadens the peaks and reduces 
their amplitude, marking the onset of attenuation-dominated 
behavior. For intermediate decay lengths, $l_c=2$ (Fig.~\ref{s88}(d)) the spectrum develops 
broad minima and two separated maxima, showing that only selected mode windows retain high 
transmission. At $l_c=3$ (Fig.~\ref{s88}(e)) the transmission becomes U-shaped with a single 
pronounced high-frequency maximum and a strongly suppressed mid-frequency band, indicating that 
longer-range coupling permits a small set of high-frequency modes to propagate efficiently while 
intermediate modes are attenuated. Finally, for $l_c=4$ (Fig.~\ref{s88}(f)) the low- and 
mid-frequency transmissions remain strongly suppressed and a sharp high-frequency peak 
appears close to the silicon cut-off, consistent with transport dominated by long-wavelength, 
coherent phonons. 

Overall, the results reveal a 
clear crossover from an interference-dominated regime at small $l_c$ to a regime governed 
primarily by decay and long-range coupling at larger $l_c$.

Figure~\ref{s8} presents the phonon transmission probability $\mathcal{T}_{\text{ph}}$ for the 
helical geometry with germanium leads for various values of the decay length $\ell_c$. 
Although the overall interference-to-attenuation crossover is similar to the silicon case, 
the spectral features differ markedly due to the heavier atomic mass of Ge and its 
significantly smaller phonon bandwidth. As a consequence, the transmission spectra exhibit 
fewer oscillations, broader resonances, and a more rapid suppression of high-frequency 
components.
\begin{figure}[H]
\includegraphics[width=0.33\textwidth]{trans_ph_lc0.1.eps}\hfill\includegraphics[width=0.33\textwidth]{trans_ph_lc0.5.eps}\hfill\includegraphics[width=0.33\textwidth]{trans_ph_lc1.eps}\vskip 0.1 in
\includegraphics[width=0.33\textwidth]{trans_ph_lc2.eps}\hfill\includegraphics[width=0.33\textwidth]{trans_ph_lc3.eps}\hfill\includegraphics[width=0.33\textwidth]{trans_ph_lc4.eps}
\caption{(Color online). Phonon transmission probability $\mathcal{T}_{\text{ph}}$ as a function of phonon angular frequency $\omega$ for germanium leads. (a) $l_c=0.1\,$\AA, (b) $l_c=0.5\,$\AA, (c) $l_c=1\,$\AA, (d) $l_c=2\,$\AA, (e) $l_c=3\,$\AA, and (f) $l_c=4\,$\AA. The temperature is fixed at $T=300\,$K.}
\label{s8}
\end{figure} 
For the shortest decay length, $\ell_c = 0.1$ (Fig.~\ref{s8}(a)), Fabry--P\'{e}rot--like 
oscillations appear across the spectrum, but the number of peaks is considerably smaller 
than in silicon, reflecting the narrower Ge phonon bandwidth and reduced contribution of 
high-frequency modes. Increasing the decay length to $\ell_c = 0.5$ 
(Fig.~\ref{s8}(b)) smoothens the oscillatory pattern, leaving only a few well-separated 
resonant structures; in contrast, silicon still exhibits dense oscillations in this regime.
At $\ell_c = 1$ (Fig.~\ref{s8}(c)), the transmission develops a broad valley at intermediate 
frequencies followed by one or two prominent peaks before sharply dropping at the Ge 
cut-off. The number of resonance--antiresonance pairs remains substantially fewer than in 
the corresponding silicon spectra. As the decay length increases to $\ell_c = 2$ 
(Fig.~\ref{s8}(d)), interference features weaken significantly, yielding a smooth 
valley--peak profile indicative of growing attenuation.
For larger decay lengths, $\ell_c = 3$ and $\ell_c = 4$ (Figs.~\ref{s8}(e)--(f)), the 
transmission becomes almost completely monotonic, with only a mild upturn near the 
high-frequency edge for $\ell_c = 3$ and an essentially featureless decay for 
$\ell_c = 4$. This monotonic behaviour appears much earlier than in silicon, underscoring 
the reduced high-frequency phonon mode density of germanium.

Overall, germanium leads exhibit the same qualitative evolution from interference-dominated 
to attenuation-dominated transport as $\ell_c$ increases, but with markedly fewer peaks, weaker 
oscillatory contrast, and a more restricted frequency range compared to silicon. These 
differences stem from the heavier mass of Ge atoms and their correspondingly smaller Debye 
frequency, which limit the available vibrational channels and suppress high-frequency 
interference effects.

In Fig.~\ref{s9}, we present the variation of the phonon thermal conductance $k_{\rm ph}$ as a function of the decay constant $l_c$ for silicon and germanium leads. For silicon (Fig.~\ref{s9}(a)), $k_{\rm ph}$ exhibits a pronounced oscillatory behavior with a gradually decreasing envelope as $l_c$ increases. The strong oscillation amplitudes indicate that the phonon thermal conductance is highly sensitive to the decay constant $l_c$. Within the considered $l_c$-range, $k_{\rm ph}$ varies from as high as $70\,$pW/K to as low as $10\,$pW/K. These relatively large conductance values stem from the wider phonon bandwidth of silicon, arising from its higher cut-off frequency and lighter atomic mass, which together produce stronger mode mixing and interference.
\begin{figure}[h]
\includegraphics[width=0.45\textwidth]{lc_var_si.eps}\hfill\includegraphics[width=0.45\textwidth]{lc_var.eps}
\caption{(Color online.) Behavior of thermal conductance due to phonons $k_{\text{ph}}$ as a function of the decay constant $l_c$ for (a) silicon and (b) germanium leads. The dotted line is just for the aid to the eye. The temperature is fixed at $T=300\,$K.}
\label{s9}
\end{figure} 
In contrast, the germanium case (Fig.~\ref{s9}(b)) displays a similar oscillatory pattern but with substantially reduced amplitude, with conductance values mostly spanning from about $40\,$pW/K down to only a few pW/K. This reduction reflects the heavier atomic mass of germanium and its narrower phonon spectrum, both of which restrict the number of available propagating vibrational modes. Overall, silicon leads support a higher and more strongly modulated phononic heat transport compared to germanium, consistent with the intrinsic differences in their phonon dispersion characteristics.

\end{document}